\documentstyle[sprocl,epsfig]{article}

\def\Journal#1#2#3#4{{#1} {\bf #2}, #3 (#4)}

\def\NIMA{{\em Nucl. Instrum. Methods} A}

\def\PRL{\em Phys. Rev. Lett.}
\def\PRD{{\em Phys. Rev.} D}
\def\PRC{{\em Phys. Rev.} C}
\def\ZPC{{\em Z. Phys.} C}

\def\be{\begin{equation}}
\def\ee{\end{equation}}
\def\bea{\begin{eqnarray}}
\def\eea{\end{eqnarray}}

\begin{document}

\vspace{-.2cm}
\title{Hadron Formation in DIS in a nuclear environment}
\vspace{-2.cm}
\author{Valeria Muccifora (on behalf of the HERMES Collaboration)}
\address{I.N.F.N. 
 Laboratori Nazionali di Frascati, via E. Fermi 40 I-00044 Frascati, Italy\\
E-mail: valeria.muccifora@lnf.infn.it} 
\vspace{-.4cm}
\maketitle\abstracts{The influence of the nuclear medium on the production of charged hadrons in semi-inclusive deep inelastic scattering has been studied by the HERMES experiment at DESY using 27.5 GeV positrons. A substantial reduction of the multiplicity of charged hadrons and identified charged pions from nuclei relative to that from deuterium has been measured as function of the relevant kinematic variables.
 The preliminary results on krypton show a larger reduction of the multiplicity ratio $R_M^{h}$ with respect to the one previously  measured on nitrogen
 and suggest a possible modification of the quark fragmentation process in the nuclear environment.}
\vspace{-1.cm}
\footnote{Proceeding of DIS 2001, IX International Workshop on Deep Inelastic Scattering\\
Bologna, 27 April-1 May 2001}
\section{Introduction}
 The understanding of quark propagation through the nuclear environment is crucial for the interpretation  of  ultra-relativistic heavy ion collisions and 
 high energy proton-nucleus  and  lepton-nucleus interactions. Quark propagation in the nuclear medium involves competing processes like
the  hadronization of quarks, the quark energy loss through multiple scattering, and gluon radiation. 
Semi-inclusive deep inelastic lepton-nucleus collisions provide  a unique opportunity to study these effects. In the simplest scenario, the nucleus, which has the size of a few fermi, acts as an ensemble of targets with which the struck quark or the 
formed hadron may interact. In contrast to proton-nucleus scattering, in  deep inelastic scattering (DIS) no deconvolution of the distributions of the projectile and target fragmentation particles has to be made, so that hadron distributions and multiplicities from different nuclei can be
directly related to nuclear effects in quark propagation and hadronization.
\vspace{-.2cm}
\section{Experimental results} 
 The experimental results are presented in terms of the 
multiplicity ratio $R_M^{h}$, which 
represents the ratio of the number of hadrons of type $h$
produced per DIS event for 
a nuclear target of mass A to that from a deuterium target (D):
\vspace{-.2cm}
\begin{equation}
 R_M^{h}(z,\nu) = {\frac {\left. \frac{N_h(z,\nu)}{N_e(\nu)}\right|_A }
			{\left. \frac{N_h(z,\nu)}{N_e(\nu)}\right|_D }}
=  \frac{\left. \frac{\sum e^2_f q_f(x) D^h_f(z)}{\sum e^2_f q_f(x)}\right|_A}
         {\left. \frac{\sum e^2_f q_f(x) D^h_f(z)}{\sum e^2_f q_f(x)}\right|_D}.
\end{equation}
\noindent
\vspace{-.1cm}
Here
$z$ represents the fraction of the virtual photon energy $\nu$ transferred to the hadron,
 $N_h(z,\nu)$  the number of semi-inclusive hadrons in a 
given ($z,\nu$)-bin, 
$N_e(\nu)$ the number of inclusive DIS leptons in the same $\nu$-bin.
 This ratio can be expressed in term of  the fragmentation functions 
$D^h_f(z)$   of a quark of flavor $f$,
and  the quark distribution functions $q_f(x)$, with $x$ the Bjorken scaling variable.
\begin{figure}[h]
\vspace{-.2cm}
\hspace{2.cm}
    \epsfig{figure=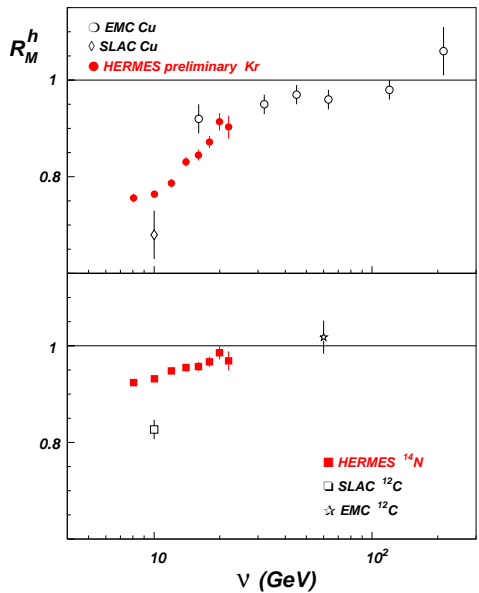,width=6.2cm}
\vspace{-.3cm}
\caption{Charged hadron multiplicity ratio $R_M^{h}$
as a function of $\nu$ for values of $z$ larger
than 0.2. 
 The error bars
represent the statistical uncertainty only.}
\label{fig:fig1}
\end{figure}

\begin{figure}[h]
\vspace{-.3cm} 
 \begin{tabular}{c c}
	\epsfig{figure=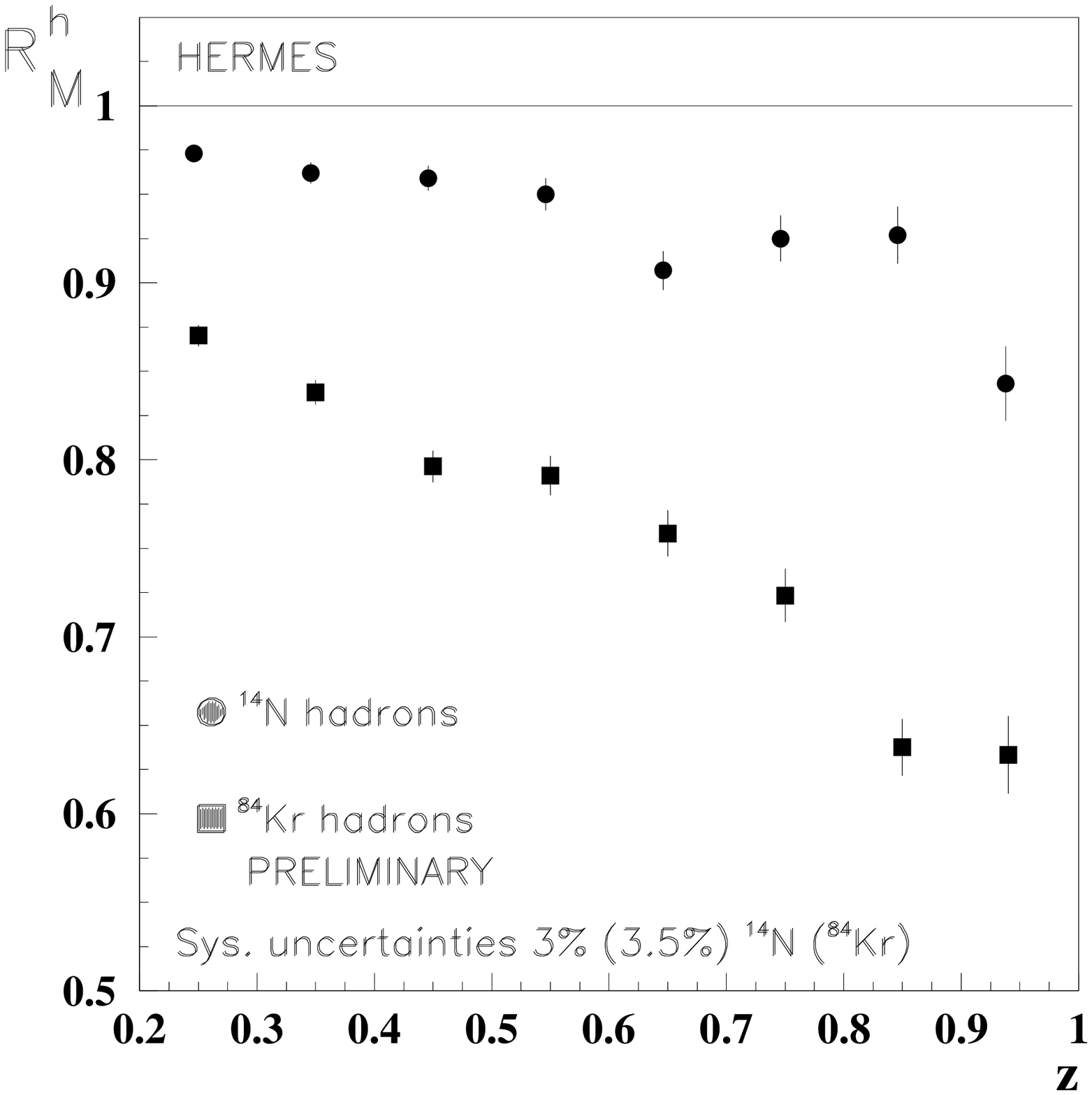,width=5.5cm}
	\epsfig{figure=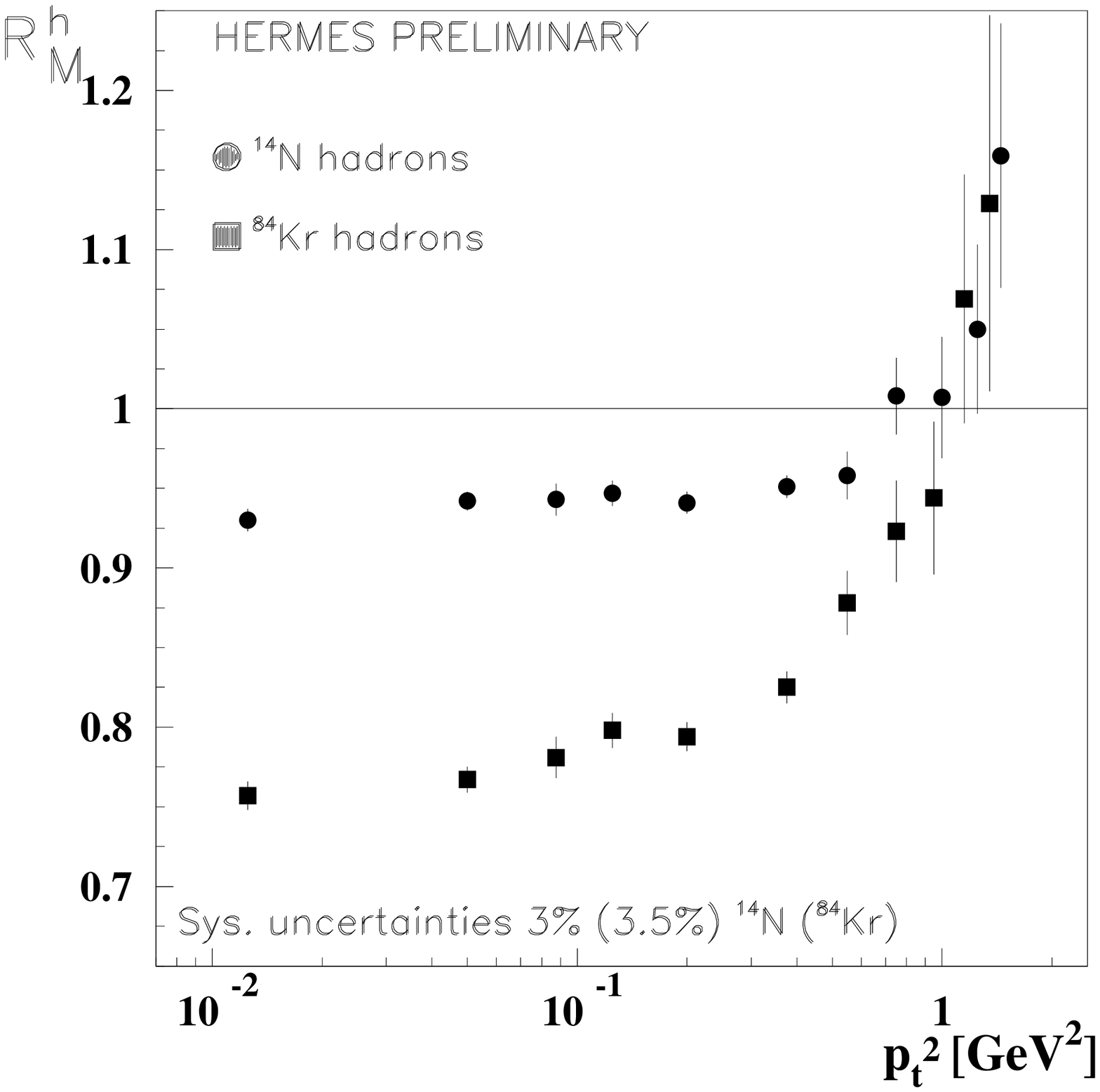,width=5.5cm}
   \end{tabular}
\vspace{-.3cm} 
\caption{The multiplicity ratio for charged hadrons  (left)  versus $z$ and (right) versus $p_t^2$.}
\label{fig:fig2}
\end{figure}

The multiplicity ratio has been measured at HERMES
using D, $^{14}$N and  $^{84}$Kr  gas targets internal to the 
 27.5 GeV HERA positron storage ring,  by identifying both the scattered 
positron and the produced hadrons in the HERMES spectrometer~\cite{AK98}.
The HERMES results~\cite{AIR01}  for the multiplicity ratio for
all charged hadrons with $z>$ 0.2
are presented as a function of $\nu$
in Fig. \ref{fig:fig1} together with
data of previous experiments on nuclei of similar size~\cite{osborne,EMC}. 
The HERMES data for $R_M^{h}$ are observed to increase with
increasing $\nu$ and are 
consistent with the high-energy EMC data.
In particular, the preliminary $^{84}$Kr data show that the HERMES energy range is well suited for the study of quark propagation and hadronization.
The energy dependence of the data follows the expectations of the gluon-bremsstrahlung model of hadronization \cite{boris,AIR01}. 
A stronger attenuation is observed for  $^{84}$Kr with respect to 
$^{14}$N   with an increase in the average attenuation of a factor of about 3.7. This value is in reasonable agreement with the  predicted modifications  of the  quark fragmentation functions in DIS~\cite{GUO01}, that are expected to  depend quadratically on the nuclear size.
The  behaviour of the  $^{14}$N  and the $^{84}$Kr  data is consistent when plotted  as a function of the relevant kinematic variables,
as  shown in Fig. \ref{fig:fig2} where  
 the multiplicity ratio for charged hadrons with $\nu >$ 7 GeV is given as a function of $z$ and  of the hadron transverse momentum square $p_t^2$.
The stronger decrease with $z$ observed for the $^{84}$Kr with respect to the 
$^{14}$N follows the expectations of the medium-modified fragmentation function~\cite{GUO01}.
 \vspace{-.2cm}    
A nuclear enhancement at high $p_t^2$ is observed from the HERMES data, similar to the one reported for proton-nucleus and nucleus-nucleus collisions which is known as the Cronin effect. The Cronin effect has been explained in the framework of multiple parton scattering. Within the Glauber formalism~\cite{WA01}  the transition  between
soft and hard processes is predicted to occur at  a scale of  $p_t$~$\sim$ 1-2 GeV in agreement with  the data of Fig. \ref{fig:fig2}.
The multiple scattering process is directly associated with multiparton correlation functions. It has been  shown~\cite{GUO02} that the transverse momentum broadening of leading pions in deep inelastic lepton-nucleus scattering is an excellent observable to probe the parton correlation functions in the nucleus, and that a measurement of 
the dependence
 of the transverse momentum enhancement  provides information on the functional form of the parton correlation functions.
The HERMES results shown in Fig. \ref{fig:fig2} seem to suggest a  nuclear-size dependence of the transverse momentum enhancement. 
The results presented thusfar concern the sum of positive and negative
hadrons. For both hadrons and pions the multiplicity ratios  have been 
separately determined for the two charge states. 
 \begin{figure}[h]
\vspace{-.3cm}
 \begin{tabular}{c c}
	\epsfig{figure=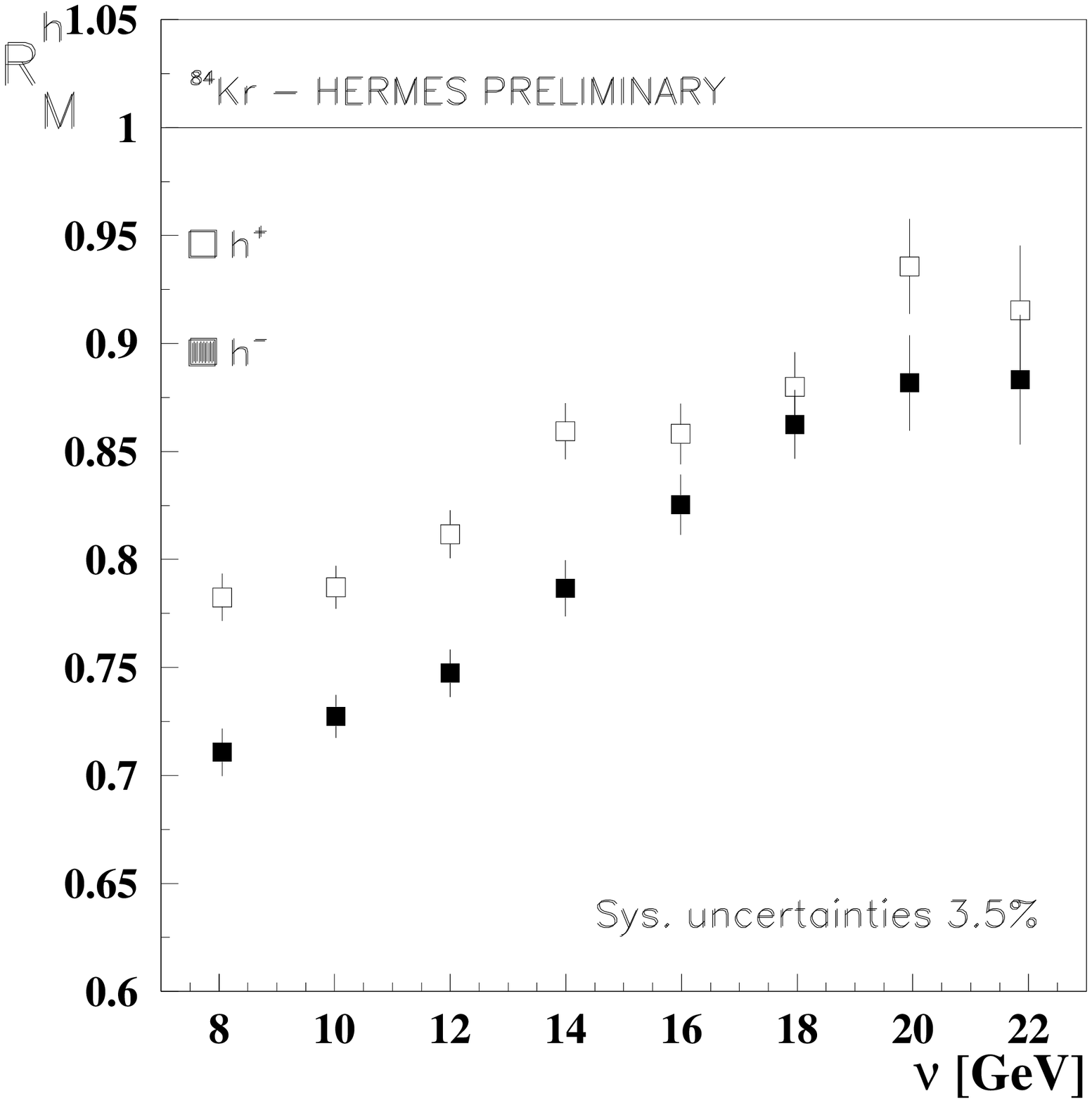,width=5.5cm}
    \epsfig{figure=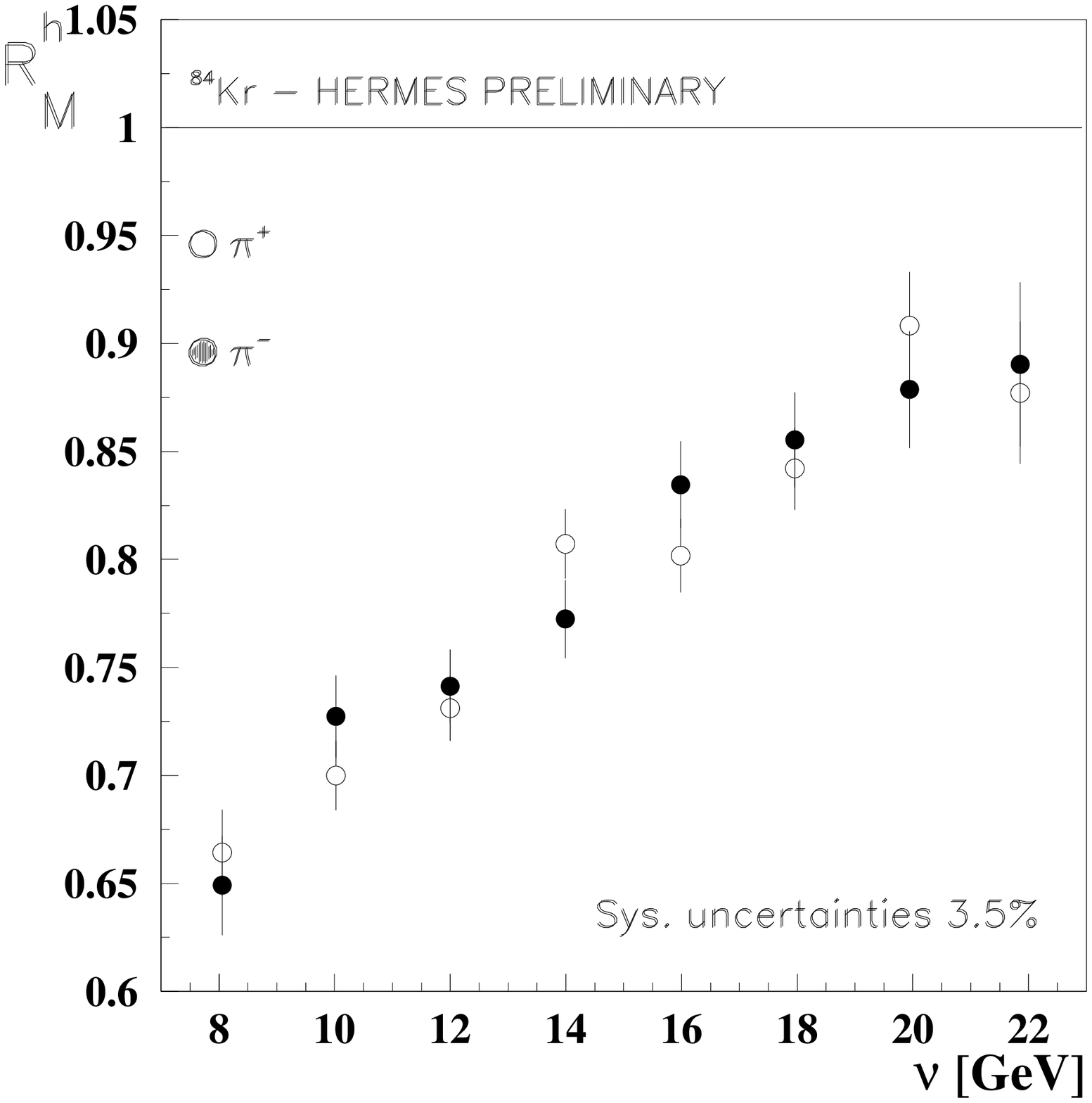,width=5.5cm}
   \end{tabular}
\vspace{-.4cm}
\caption{Multiplicity ratios for hadrons including pions (left) and
	identified pions (right) as a function of $\nu$. The
	open (closed) symbols represent the positive (negative)
	charge states.}
\label{fig:fig3}
\end{figure}
Pions were identified in the momentum range between 4 and 13.5 GeV
 by using a RICH detector. 
In the left (right) panel of Fig. \ref{fig:fig3}, the 
multiplicity ratios for positive and negative
hadrons (pions) are displayed as a function of $\nu$ for $^{84}$Kr. 
The data show that the multiplicity ratio  is the same for
positive and negative pions while a significant difference is observed between $R_M^{h}$ for
positive and negative hadrons.
This result, that agrees with the one reported for the  $^{14}$N data~\cite{AIR01}  obtained with a threshold \v Cerenkov detector,
can be interpreted in terms of a difference between the formation time of 
protons and pions. 
Alternatively, it has been  suggested that the observed  differences between positive and negative hadrons
can  be attributed to  a different modification of the quark and antiquark fragmentation
functions in nuclei~\cite{GUO04}.
In order to clarify this issue an analysis  has been started of multiplicity ratios for identified kaons, protons and anti-protons in various nuclei using the RICH detector  at HERMES.

\end{document}